\documentclass[aps,prl,10pt,twocolumn,showpacs,superscriptaddress]{revtex4-1}
\usepackage{graphicx}
\usepackage{enumitem}
\def\go{\mathcal{G}_0}
\def\gzero{$\mathcal{G}(E=0)$}

\begin{document}

\title{
Many-body correlations and coupling in benzene-dithiol junctions
}

\author{T. Rangel}
\altaffiliation[Present address: ]{Molecular Foundry, Lawrence Berkeley National Laboratory, Berkeley, CA 94720, USA.}
\affiliation{Institute of Condensed Matter and Nanosciences (IMCN), Universit\'e Catholique de Louvain, Chemin des \'Etoiles 8, bte L7.03.01, 1348 Louvain-la-Neuve, Belgium}
\affiliation{European Theoretical Spectroscopy Facility (ETSF)}
\author{A. Ferretti}
\affiliation{CNR, Istituto Nanoscienze, S3 Center, 41125 Modena, Italy}
\affiliation{European Theoretical Spectroscopy Facility (ETSF)}
\author{V. Olevano}
\affiliation{Universit\'e Grenoble Alpes, 38000 Grenoble, France}
\affiliation{CNRS, Institut N\'eel, 38042 Grenoble, France}
\affiliation{European Theoretical Spectroscopy Facility (ETSF)}
\author{G.-M. Rignanese}
\affiliation{Institute of Condensed Matter and Nanosciences (IMCN), Universit\'e Catholique de Louvain, Chemin des \'Etoiles 8, bte L7.03.01, 1348 Louvain-la-Neuve, Belgium}
\affiliation{European Theoretical Spectroscopy Facility (ETSF)}

\date{\today}

\begin{abstract}
Most theoretical studies of nanoscale transport in molecular junctions rely on the combination of the Landauer formalism with Kohn-Sham density functional theory (DFT) using standard local and semilocal functionals to approximate exchange and correlation effects.
In many cases, the resulting conductance is overestimated with respect to experiments.
Recent works have demonstrated that this discrepancy may be reduced when including many-body corrections on top of DFT.
Here we study benzene-dithiol~(BDT) gold junctions and analyze the effect of many-body perturbation theory (MBPT) on the calculation of the conductance with respect to different bonding geometries.
We find that the many-body corrections to the conductance strongly depend on the metal-molecule coupling strength.
In the BDT junction with the lowest coupling, many-body corrections reduce the overestimation on the conductance to a factor 2, improving the agreement with experiments.
In contrast, in the strongest coupling cases, many-body corrections on the conductance are found to be sensibly smaller and standard DFT reveals a valid approach.
\end{abstract}

%\pacs{85.65.+h,71.10.-w,72.10.-d,73.40.-c,73.63.-b}
%\keywords{Benzene-diamine,GW,transport,junction,correlation}

\maketitle

%%%%%%%%%%%%%%%%%%%%%%
\section{Introduction}
%%%%%%%%%%%%%%%%%%%%%%
Among the various physical observables in molecular electronic devices, the lead-molecule-lead conductance [most particularly, its zero bias value \gzero] is one of the most important characteristics.
After many years of uncertainty, reliable experimental measures of \gzero\ in molecular junctions have recently become available \cite{ venkataraman_dependence_2006}.
Today's standard first-principles approach to calculate quantum conductance relies on a combination of Kohn-Sham density-functional theory (DFT) and Landauer formalism \cite{ DiVentra, Nardelli1999, choi_ihm99prb, di_ventra_first-principles_2000, brandbyge_2002prb}.
Though in principle without direct physical meaning \cite{onida2002rmp, gatti_understanding_2007}, the DFT electronic structure of the molecular device is often used as an approximation to the quasiparticle electronic structure into the Landauer formula for the conductance. 
It was found that the most popular DFT exchange-correlation approximations, the local-density approximation (LDA) and the generalized gradient approximation (GGA), can overestimate sensibly the zero-bias conductance, as compared to the experiments \cite{lindsay_molecular_2007, di_ventra_comment_2009, di_ventra_first-principles_2000}.

Various investigations have been conducted in order to analyze the possible causes of this problem \cite{ mera_assessing_2010, sai_2005prl, kurth_2005prb, vignale_diventra_2009prb, koentopp_2006prb}, and a number of practical solutions have been proposed.
On the one hand, generalized Kohn-Sham DFT relying on full and range separated hybrid functionals \cite{ferretti_textitab_2012, BillerKronik11, tamblyn_2014jpcl, Egger_2015nanolett}, DFT+$U$ \cite{ cococcioni_degironcoli_2005, himmetoglu_2014ijqc, calzolari_2007nanotech, sclauzero_2013prb, timoshevskli_2014jpcm}, and self-interaction corrections \cite{ toher_effects_2008, filippetti2011,dabo2013pccp, nguyen2015prl} have been suggested to provide a better description of the underlying electronic structure.
On the other hand, other approaches \cite{QuekNeaton07, MowbrayThygesen08, QuekNeaton09, QuekVenkataraman09, WidawskyVenkataraman12, Egger_2015nanolett}, often referred to as DFT+$\Sigma$, correct the DFT electronic structure by a classical image-charge self-energy model parametrized on the quasiparticle corrections in the isolated molecule and the molecule-metal surface distance.
Alternatively, many-body perturbation theory (MBPT) is in principle an exact framework to calculate the electronic structure.
Within MBPT, the $GW$ approximation \cite{hedin_new_1965} has been proposed to describe the electronic structure and conductance more accurately than DFT \cite{DarancetOlevano07, ThygesenRubio08, RangelRignanese11, StrangeThygesen11, JinThygesen14}.
Both the DFT+$\Sigma$ approach and MBPT \textit{ab initio} $GW$ reduce the conductance and improve the agreement with experiments.
In the most intuitive explanation, the many-body correction to the DFT HOMO-LUMO energy gap, taking into account also the screening of the leads, is the key to reduce the conductance \cite{QuekNeaton07, MowbrayThygesen08}.
In a more complex scenario, corrections to orbital energies are not sufficient and many-body corrections to DFT wave functions are required to improve the calculation of the conductance \cite{RangelRignanese11}.

The purpose of this work is to clarify the effect of many-body corrections on DFT in the frame of quantum transport simulations and to better understand the physical mechanisms responsible for the reduction of the conductance towards improved agreement with experiments.
To this aim, we focus on the benzene-dithiol (BDT) molecule connected to gold leads (BDT@Au).
This is one of the most studied systems, both a prototype for experiments and a benchmark for theory. 
Here, we consider three different BDT@Au bonding geometries and investigate the relation between the latter (together with the resulting metal-molecule coupling) and many-body corrections to the quantum conductance.

We find that these corrections strongly depend on the bond strength and the metal-molecule coupling regime.
This dependence cannot be described by a simplified DFT+$\Sigma$ image-charge model \cite{QuekNeaton07}.
Indeed only two ingredients enter this model: the correction to the DFT underestimation of the HOMO-LUMO gap for the isolated molecule; the reduction of the correction due to the screening of the leads, as accounted by a classical image-charge model.
For the three junctions studied here, both ingredients do not change: the molecule is the same (BDT) and the metal-molecule distance is very similar, so the model cannot account for the variations in the conductance correction among the three different geometries.
Finally, in the limit of strong coupling, many-body corrections have been found to have a negligible effect on the conductance such that DFT reveals a good approximation in these cases.

\begin{figure}%[t]
\includegraphics[width=\columnwidth]{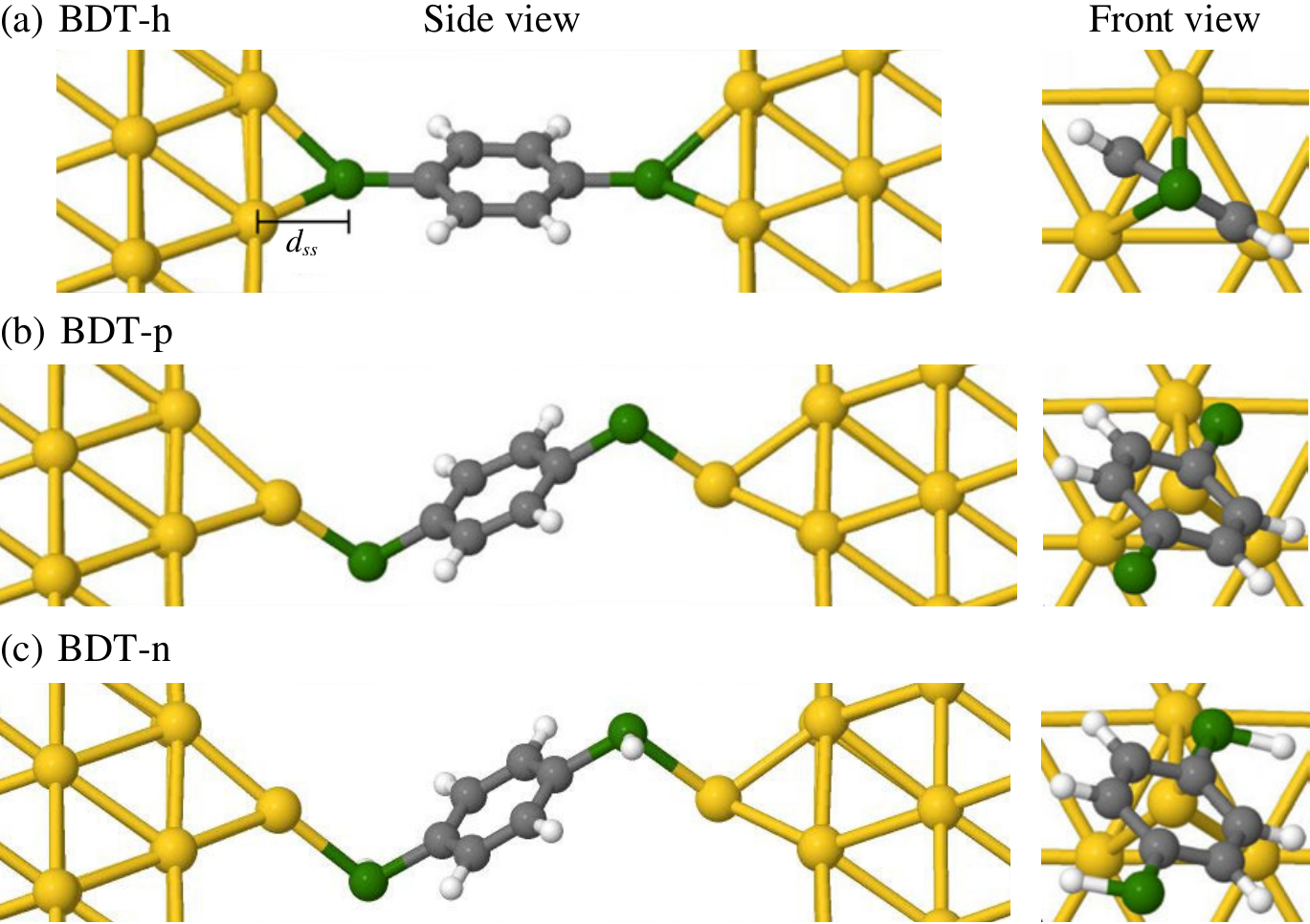}
\caption[BDT geometries]{
Side and front views for the different geometries of BDT attached to gold (111):
(a) BDT-h, (b) BDT-p, and (c) BDT-n.
For sake of clarity, all gold layers are not shown in the figures: only one (respectively 4) in the front (respectively side) views.
The Au, C, S, and H atoms are represented by yellow (light grey), grey, green (dark grey), and white spheres, respectively.
}
\label{fig:junctions-geom}
\end{figure}

%%%%%%%%%%%%%%%%%%%%%%%%%%%%%%%%%%%%%
\section{BDT@Au junctions geometries}
%%%%%%%%%%%%%%%%%%%%%%%%%%%%%%%%%%%%%
In experiments measuring the molecular conductance, the atomic structure of the molecular junction is often not well characterized.
Due to this experimental uncertainty, \textit{ab initio} calculations have been used to suggest several possible geometries and characterize their stability.
We have considered three geometries (see Fig.~\ref{fig:junctions-geom}) for BDT attached to Au (111) leads, we explain their relevance in the next paragraphs:
(i) in the {\it hollow geometry}~(BDT-h), each S atom loses its H atom and binds to three Au atoms of the flat lead surface (at about the same distance);
(ii) in the {\it pyramid geometry}~(BDT-p), each S atom loses its H atom and binds to a Au adatom forming a pyramid on the surface;
(iii) in the {\it nondissociative geometry} (BDT-n), each S atom binds to a Au adatom as in BDT-p but without losing its H atom.

Initially, the BDT-h geometry was considered to be the preferred bonding site for the thiol radical \cite{ sellers_structure_1993}.
Later studies pointed out different possible adsorption geometries \cite{ hakkinen_electronic_1999, gronbeck_thiols_2000, hayashi_adsorption_2001, kruger_interaction_2001, yourdshahyan_$n$-alkyl_2001}.
A comparison of binding energies revealed that BDT-p is more stable than BDT-h by $\sim$0.4~eV per molecule \cite{ning_first-principles_2007}.
More recently, theory and experiments found that the not-dissociative geometry (BDT-n) is possible and energetically favorable \cite{zhou_methanethiol_2006, ning_first-principles_2007, demir_identification_2012}.

Since our purpose is to study the effect of many-body corrections on the conductance as a function of the hybridization and the bonding geometry, we consider all three geometries here.
In fact, already at the DFT level, the conductance was found to strongly depend on the junction structure \cite{ stokbro_theoretical_2003, tomfohr_theoretical_2004, muller_effect_2006}.

The relaxed geometries used in this work are shown in Fig.~\ref{fig:junctions-geom}.
They agree with previous calculations \cite{MowbrayThygesen08, ning_first-principles_2007, toher_effects_2008, ning_correlation_2014}.
The tilt angle between the normal to the gold surface and the S-S direction of the two contact atoms measures $32^\circ$, $31^\circ$, and $0^\circ$ for BDT-n, BDT-p and BDT-h, respectively.
Moreover, the Au-S distance is~2.5, 2.3, and 2.6~\AA\ for BDT-n, BDT-p and BDT-h, respectively.
For BDT-h, the distance $d_{ss}$ between the S atom and the first Au layer, as shown in Fig.~\ref{fig:junctions-geom}, is 2.0~\AA.

%%%%%%%%%%%%%%%%
\section{Theory}
%%%%%%%%%%%%%%%%
\subsection{Landauer Fisher-Lee conductance formula}

In the Green-function formalism, the conductance of the junction is provided by the Landauer formula as in the Fisher-Lee form (in units of the conductance quantum ${\cal G}_0 = 2 e^2 / h$):
\begin{equation}
{\cal G}(E) =  \mathop{\mathrm{tr}} \left[\Gamma_L(E) G_C^r(E) \Gamma_R(E) G_C^a(E) \right].
\label{eq:landauer}
\end{equation}
where $G_C^a$ and $G_C^r$ are the advanced and retarded Green's functions for the central region and $\Gamma_L$ and $\Gamma_R$ are the coupling to the left and right leads.
All these quantities can be calculated from the Green's functions of the uncoupled systems [$\alpha = \mathrm{L}$, R (left and right leads), and C (central region)] which are related to their electronic structures (both energies $\epsilon_{\alpha,i}$ and wave functions $\phi_{\alpha,i}$):
\begin{equation}
g_\alpha^{a/r}(\mathbf{r},\mathbf{r}',E) = \sum_i \frac{\phi_{\alpha,i}(\mathbf{r}) \phi^*_{\alpha,i}(\mathbf{r}')}{E - \epsilon_{\alpha,i} \pm i \eta}.
\label{green}
\end{equation}

Today's standard first-principles approach consists of using the electronic structure obtained from DFT.
In this work, the Landauer conductance is calculated not only using this standard approach but also adopting three different electronic structures based on MBPT, as explained in the following sections.

\subsection{Many-body perturbation theory vs DFT}

In many-body perturbation theory (MBPT) the quasiparticle (QP) electronic structure [i.e., both energies $\epsilon_i^\mathrm{QP}$ and wave functions $\phi_i^\mathrm{QP}(r)$] is calculated by solving the quasiparticle equation,
\begin{eqnarray}
H_0(\mathbf{r}) \phi_i^\mathrm{QP}(\mathbf{r}) + \int d\mathbf{r}' \, 
\Sigma(\mathbf{r},\mathbf{r}',\omega=\epsilon_i^\mathrm{QP}) \, \phi_i^\mathrm{QP}(\mathbf{r}')
\nonumber \\
= \epsilon_i^\mathrm{QP} \phi_i^\mathrm{QP}(\mathbf{r}) \, , \hspace{2cm}
\label{qp-eq}
\end{eqnarray}
where $H_0(\mathbf{r}) = -\nabla^2/2 + v_\mathrm{ext}(\mathbf{r}) + v_\mathrm{H}(\mathbf{r})$ is the local Hamiltonian containing the kinetic term, the interaction with the external potential $v_\mathrm{ext}(\mathbf{r})$, and the Hartree term $v_\mathrm{H}(\mathbf{r})$.
Many-body exchange and correlation (XC) effects (beyond the Hartree approximation) are taken into account through the self-energy $\Sigma(\mathbf{r},\mathbf{r}',\omega)$ operator which is non-Hermitian, nonlocal, and frequency dependent.

The calculation of the exact self-energy is unfeasible for real complex systems, so that one must resort to approximations. 
By comparing Eq.~(\ref{qp-eq}) with the DFT Kohn-Sham equation:
\begin{equation}
  \left[ H_0(\mathbf{r}) + v_\mathrm{xc}(\mathbf{r}) \right] \phi_i^\mathrm{DFT}(\mathbf{r})
  = \epsilon_i^\mathrm{DFT} \phi_i^\mathrm{DFT}(\mathbf{r})
  ,
\label{ks-eq}
\end{equation}
the exchange-correlation potential $v_\mathrm{xc}(\mathbf{r})$ can be seen as a mean-field, local, static, and Hermitian approximation to the self-energy $\Sigma(\mathbf{r},\mathbf{r}',\omega)$.
This is the rationale for the use of the DFT electronic structure, $\epsilon_i^\mathrm{DFT}$ and $\phi_i^\mathrm{DFT}(\mathbf{r})$, as an approximation to the true QP electronic structure.

It is however worth pointing that, in the DFT formalism, $v_\mathrm{xc}(\mathbf{r})$ is \textit{not} introduced as an approximation to $\Sigma(\mathbf{r},\mathbf{r}',\omega)$.
In fact, the DFT theorems guarantee the existence of a local $v_\mathrm{xc}(\mathbf{r})$ which provides the exact electronic density $\rho(\mathbf{r})$ and the exact ground-state energy $E$ of the real (i.e., including many-body effects) system.
But DFT theorems and the exact $v_\mathrm{xc}(\mathbf{r})$ are not concerned with the electronic structure, quasiparticle energies $\epsilon_i^\mathrm{QP}$ and wave functions $\phi_i^\mathrm{QP}$, of the real system.
Then, since the exact form of $v_\mathrm{xc}(\mathbf{r})$ is unknown, it needs to be approximated using local or semilocal XC functionals, e.g., LDA, GGA, etc.
%Approximated local or semi-local
Such functionals lack fundamental characteristics like the derivative discontinuity \cite{perdew_density-functional_1982} which has been shown to be essential to reproduce quantum transport correlation effects like the 0-temperature conductance plateau in the Anderson model at the weak-coupling limit \cite{stefanucci_towards_2011,bergfield_bethe_2012,troster_transport_2012}.
Furthermore, the DFT eigenvalues are in fact Lagrange multipliers which are introduced in order to ensure the normalization of the wave functions.
Formally, the DFT electronic structure has thus no physical interpretation (with the notable exception of the highest occupied state, which yields the exact ionisation potential of the system \cite{AlmbladhVonBarth85}; see also Ref.~\cite{ChongBaerends02} for the relationship between other occupied Kohn-Sham energies and vertical ionization potentials).
Nevertheless, it is reasonable to consider it as a convenient 0-order starting point for the calculation of the QP electronic structure. 

\subsection{$GW$ vs COHSEX approximations}

In the $GW$ approximation \cite{ hedin_new_1965}, the self-energy $\Sigma(\mathbf{r},\mathbf{r}',\omega)$ is obtained as the convolution of the Green function $G$ and the dynamically screened interaction $W$,
\begin{equation}
  \Sigma(\mathbf{r},\mathbf{r}',\omega) = \frac{i}{2\pi} \int d\omega' \, G(\mathbf{r},\mathbf{r}',\omega-\omega') W(\mathbf{r},\mathbf{r}',\omega')
  .
  \label{gw-eq}
\end{equation}

In principle $G$ and $W$ depend on the final QP electronic structure, so that they should be calculated self-consistently solving iteratively a closed set of integrodifferential equations known as Hedin equations \cite{hedin_new_1965}.
This approach is referred to as the self-consistent $GW$ approximation.
There are still many open questions regarding such calculations for real systems, but also for the simple jellium model \cite{HolmVonBarth}, in particular with respect to their effective improvement of the electronic structure compared to non-self-consistent calculations \cite{HolmVonBarth}.
%Our aim here is clearly not address this issue.
%And, due to the complexity of the investigated systems, we could not afford such calculations.
In practice, it is customary to perform one iteration, non-self-consistent $GW$ calculations, referred to as $G_0W_0$ in the literature, using the $G$ and $W$ built from the DFT electronic structure.

As an alternative to the $G_0W_0$ approach, we use the Coulomb-hole screened-exchange (COHSEX) approach \cite{ hedin_new_1965,hybertsen_electron_1986} which has been proposed as a static approximation on top of the $GW$ self-energy:
\begin{eqnarray}
& & \Sigma^\mathrm{COHSEX}(\mathbf{r},\mathbf{r}') = 
\Sigma^\mathrm{COH}(\mathbf{r},\mathbf{r}') +
\Sigma^\mathrm{SEX}(\mathbf{r},\mathbf{r}')
\nonumber\\
& & \Sigma^\mathrm{COH}(\mathbf{r},\mathbf{r}') =
\delta(\mathbf{r},\mathbf{r}')
\left[ W(\mathbf{r},\mathbf{r}',\omega=0) - v(\mathbf{r}-\mathbf{r}') \right]
\nonumber \\
& & \Sigma^\mathrm{SEX}(\mathbf{r},\mathbf{r}') =
-\sum_v \psi_v(\mathbf{r})\psi_v^*(\mathbf{r}')W(\mathbf{r},\mathbf{r}',\omega=0).
\end{eqnarray}
Since the COHSEX self-energy is $\omega$ independent and Hermitian, it represents an important simplification with respect to the $G_0W_0$ one which is very useful in order to perform a full diagonalization of the QP Hamiltonian (see the next section).

The $G_0W_0$ approximation has been demonstrated to provide an electronic structure and band gaps in good agreement with experiments for a wide variety of materials.
The COHSEX approximation contains almost all the physics of the $GW$ approximation, neglecting only the dynamical effects.
As a result, it also leads to a significant improvement over DFT.
However, COHSEX band gaps are usually larger than those from $GW$ and from experiments, due to an overcorrection of the DFT underestimation of the band gap.
However, since our aim is not to reproduce the experimental results exactly but rather to study and understand how many-body effects correct the DFT conductance, this overcorrection is not really a problem, especially in consideration of the possibility to perform a full diagonalization of the self-energy (see the next section). 

\subsection{Diagonal vs full calculations}

In practical calculations, it is very common to apply perturbation theory on top of DFT energies and wave functions, taking $\Sigma - v_\mathrm{xc}$ as a perturbation.
The first-order QP corrections to DFT energies can be calculated assuming that the QP wave functions are equal to the DFT ones as
\[
  \epsilon_i^\mathrm{QP} = \epsilon_i^\mathrm{DFT} + \langle \phi_i^\mathrm{DFT} | \Sigma(\omega=\epsilon_i^\mathrm{QP}) - v_\mathrm{xc} | \phi_i^\mathrm{DFT} \rangle
.
\]
A further complication is that this is a nonlinear equation because the energy-dependent $\Sigma(\omega=\epsilon_i^\mathrm{QP})$ has to be calculated to the quasiparticle energy unknown of the equation.
Under the assumption that the difference between QP and DFT energies is relatively small, the matrix elements of the self-energy operator can be Taylor expanded to first order around $\epsilon_i^\mathrm{DFT}$ leading to
\[
  \epsilon_i^\mathrm{QP} = \epsilon_i^\mathrm{DFT} + Z_i \langle \phi_i^\mathrm{DFT} | \Sigma(\omega=\epsilon_i^\mathrm{DFT}) - v_\mathrm{xc} | \phi_i^\mathrm{DFT} \rangle
  ,
\]
where $Z_i$ is the QP renormalization factor given by
\[
  Z_i = \left[ 1 - \bigg\langle \phi_i^\mathrm{DFT} \bigg| \left. \frac{d\Sigma}{d\omega} \right|_{\omega=\epsilon_i^\mathrm{DFT}} \bigg| \phi_i^\mathrm{DFT} \bigg\rangle \right]^{-1}
  .
\]
So, the first-order perturbation theory corrections to the energies require only the diagonal matrix elements of the self-energy.
There is no need to calculate the off-diagonal matrix elements and to perform a complete diagonalization to solve Eq.~(\ref{qp-eq}).
In this work, this diagonal approach was used both with the $G_0W_0$ and COHSEX approximations (referred to as diag-$G_0W_0$ and diag-COHSEX, respectively): only energies are corrected, while the wave functions are kept at the DFT level.

It is possible to go beyond the diagonal approaches by taking into account the off-diagonal self-energy matrix elements to calculate the first-order corrections to the wave functions \cite{Li2002}.
The same off-diagonal matrix elements also enter into the second-order corrections to the energies \cite{Kaplan2015}.
Alternatively, one can perform a full diagonalization of the QP Hamiltonian $H_0+\Sigma$ of Eq.~(\ref{qp-eq}) to get the many-body energies and wave functions correct to all orders in perturbation theory.
In this work, we decided to go directly for the full-diagonalization procedure but could only afford it within the COHSEX approximation (referred to as full-COHSEX hereafter).
Indeed, this task is much easier and more affordable within COHSEX than $G_0W_0$ given that the COHSEX Hamiltonian is Hermitian and does not depend on the frequency $\omega$.

In isolated molecules, the differences between the energies obtained by perturbation theory at the first order, at the second order, and by full diagonalization of the QP Hamiltonian have been found to be very small and within the numerical convergence error \cite{Kaplan2015}.
We expect that this should be even more so for the \textit{metallic} lead-molecule junctions considered here.
Therefore, we do not expect appreciable changes in the energies between diag-COHSEX and full-COHSEX (this cannot be checked explicitly since the diagonalization procedure breaks the one-to-one correspondence between the energies).
As a further proof, we will see in Sec.~IV~C, which is dedicated to the local density-of-states analysis, that the off-diagonal self-energy matrix elements, entering into first-order corrections to wave functions, are small.
Indeed, the difference between the full-COHSEX and DFT wave functions is small and can hardly be noticed when comparing them directly: it can only be evidenced by plotting the wave functions/local density-of-states (LDOS) difference.
This shows that perturbation theory on top of DFT with respect to $\Sigma-v_\mathrm{xc}$ clearly holds, and that the energies obtained by perturbation theory at the first order are within our numerical convergence error from those obtained by full diagonalization of the QP Hamiltonian.
Note that if this were not the case and perturbation theory on top of DFT did not hold, approaches like the DFT+$\Sigma$ image-charge model would have to be ruled out as invalid since the beginning.

\subsection{Computational details}

The ground state and MBPT calculations are performed with the \textsc{abinit} package \cite{gonze_abinit:_2009}.
The Perdew-Burke-Ernzerhof~(PBE) functional \cite{perdew_generalized_1996} is used to approximate the exchange-correlation potential.
Norm-conserving pseudopotentials are used.
For gold, these include the 5$s$ and 5$p$ semicore states which are important for the MBPT calculations \cite{rangel_band_2012}.
An energy cutoff of 30~Ha is set to expand the plane-wave basis set.
Maximally localized Wannier functions (MLWFs) for many-body quasiparticles are obtained as explained in Ref.~\cite{hamann_maximally_2009}.
The \textsc{want} package \cite{want} is used to obtain the MLWFs and to perform the transport calculations.
In this work, the parameters chosen are those used in Ref.~\cite{RangelRignanese11}.
The junctions consist of a fully relaxed $2 \times 2$ Au (111) surface with seven atomic layers in the electrodes.
For the MBPT calculations, only four layers are used.
For the BDT geometry, it was verified that using a $3 \times 3$ surface cell and
seven layers of gold does not change  significantly the conductance at the DFT and $GW$ levels.
An $8 \times 8 \times 3$ grid of $k$ points is used to sample the Brillouin zone.
The QP corrections are calculated explicitly for 210 bands at 96 irreducible $k$ points including $\sim$300 bands in the calculations.

\section{Results}

\subsection{Conductance profile}

\begin{figure}[t]
\centering
\includegraphics[width=\columnwidth]{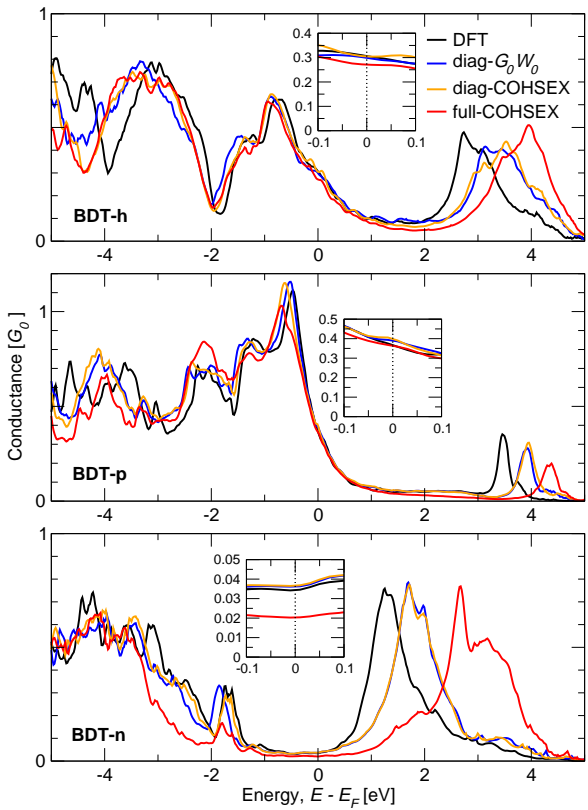}
\caption[Conductance for BDT-h, BDT-p and BDT-n]{
Conductance ${\cal G}(E)$ in units of ${\cal G}_0$ as a function of the energy $E$ for BDT-h (top panel), BDT-p (middle) and BDT-n (bottom panel).
Black line: DFT; blue line: diag-$G_0W_0$; yellow line: diag-COHSEX; red line: full-COHSEX.
The insets are zooms over the zero-bias conductance ${\cal G}(E=0)$ region.
The zero of the energy is set to the Fermi energy $E_F$.
}
\label{fig:bdtcond}
\end{figure}

The various Landauer conductances (DFT, diag-$G_0W_0$, diag-COHSEX, and full-COHSEX) as a function of energy ${\cal G}(E)$ calculated for the three different junctions are presented in Fig.~\ref{fig:bdtcond}.  
For DFT, diag-$G_0W_0$, and diag-COHSEX, the electronic structures consist of exactly the same DFT wave functions $\phi_i^\mathrm{DFT}(\mathbf{r})$ but with different energies.
In contrast, the full-COHSEX electronic structure differs from DFT both in the energies and the wave functions.

In Fig.~\ref{fig:bdtcond}, it can be seen that the diagonal approaches modify the DFT conductance profile ${\cal G}(E)$, but in none of the cases are they able to modify significantly the 0-bias conductance.
This is a confirmation of an important result that we had found previously \cite{RangelRignanese11}.

\begin{figure}[t]
\centering
\includegraphics[width=\columnwidth]{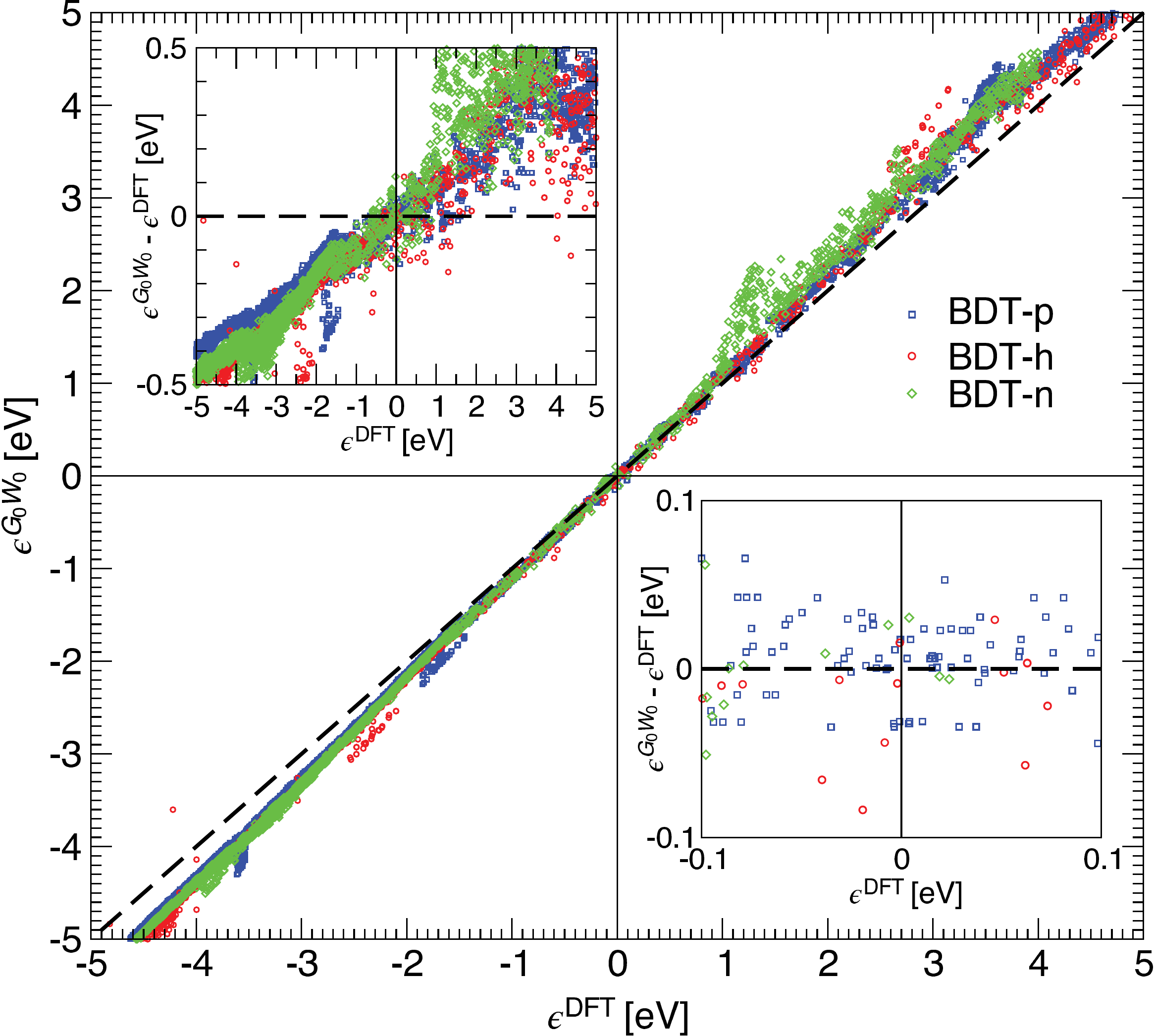}
\caption{DFT vs diagonal $G_0W_0$ energies for BDT-p (blue squares), BDT-h (red circles) and BDT-n (green diamonds).
The zero of the energy is set to the Fermi energy $E_F$.
In the insets we plot the GW corrections vs DFT energies at different scales.
}
\label{fig:gwdftenergies}
\end{figure}

In order to understand this result, we analyzed the effect of the $G_0W_0$ corrections to the DFT energies.
In Fig.~\ref{fig:gwdftenergies} we report DFT energies $\epsilon^\mathrm{DFT}$ on the $x$ axis and the corresponding $G_0W_0$ energy $\epsilon^{G_0W_0}$ on the $y$ axis for all the junction states within 5~eV from the Fermi energy.
In this figure, results lying exactly on the diagonal (which is represented as a black solid line) indicate DFT energies which coincide with the $G_0W_0$ ones (there is no effective $G_0W_0$ correction).
The corrections shown in Fig.~\ref{fig:gwdftenergies} are those typical of normal metals, with 0 or almost no correction at the Fermi energy, and an increasing correction when moving away from the Fermi level.
This is very similar to what had been found for BDA@Au in Ref.~\cite{RangelRignanese11}.
As a consequence, the Green function of the uncoupled central region $g_C^{G_0W_0}(E)$ calculated using diag-$G_0W_0$ energies and DFT wave functions in Eq.~(\ref{green}) will be different from the one calculated with both DFT energies and wave functions $g_C^{DFT}(E)$ at $E$ faraway from $E_F$, but they will be very similar at $E\simeq E_F$: $g_C^{G_0W_0}(E_F) \simeq g_C^\mathrm{DFT}(E_F)$.
Indeed, the Green function at a certain energy $E$ is mostly determined by the electronic structure around $E$, and so only by the neighboring poles, while distant poles are uninfluential (even though they undergo important shifts).

This is directly reflected in the 0-bias conductance which remains unchanged when using diag-$G_0W_0$ instead of DFT energies.
Indeed, the Landauer Fisher-Lee formula Eq.~(\ref{eq:landauer}) is also local in energy: the conductance ${\cal G}(E)$ at energy $E$ only depends on the Green function $g_\alpha(E)$ at the same energy.
Since the diag-$G_0W_0$ corrections to the DFT energies are negligible at $E_F$ (for both the central region and the leads), the changes in both the Green functions $g_\alpha(E=0)$ and the 0-bias conductance ${\cal G}(E=0)$ will also be negligible.
This is exactly what we observe in Fig.~\ref{fig:bdtcond}.
Hence, the diagonal approaches are unable to change (in particular, reduce) the DFT 0-bias conductance, although they modify the energies away from $E_F$ where the conductance profile is also altered.
In some cases (e.g., BDT-n), an increase of the 0-bias conductance can even be observed.
This is actually due to small changes, below the numerical error, in the position of the recalculated Fermi level.
In fact, in a two-level model, the zero-bias conductance was shown to depend only on the density at $E_F$ due to the Friedel sum rule \cite{mera_assessing_2010,stefanucci_towards_2011,troster_transport_2012}.
In the BDT molecular junctions considered here, there is only one eigenchannel at $E_F$. So, we expect that, for the same reason, the diagonal approaches, which do not change the eigenvalues (nor the density) at $E_F$, will have a negligible effect in \gzero.
If there are no major changes in the energies at $E_F$, the only possibility to change the 0-bias conductance is to change the wave functions at the numerator of $g_C(E)$ in Eq.~(\ref{green}).
This can only be achieved by approaches taking into account also off-diagonal self-energy matrix elements such as the full-COHSEX.

In Fig.~\ref{fig:bdtcond}, we observe that these off-diagonal elements have an important effect on the 0-bias conductance in BDT-n, some effect in BDT-h, and basically no effect in BDT-p.
This observation, which is the central finding of the present work, will be discussed in the next section.

\begin{table}[t]
\centering
\begin{ruledtabular}
\begin{tabular}{cccccc}
 & \multicolumn{2}{c}{${\mathcal{G}}$} \\
\cline{2-3}
{Junction} & DFT & full-COHSEX & ${\Delta_a \mathcal{G}}$ &
  ${\Delta_r \mathcal{G}}$     & ${E_\mathrm{coupl}}$ \\
\hline
 BDT-h & 0.305 & 0.270 & 0.035 & \phantom{.}11\% & -1.05 \\ % & -0.08 \\
 BDT-p & 0.365 & 0.362 & 0.003 &           0.8\% & -1.56 \\ % & -0.115 \\
 BDT-n & 0.034 & 0.020 & 0.014 & \phantom{.}41\% & -0.57 \\ % & -0.04 \\
\end{tabular}
\end{ruledtabular}
\caption{\label{table:cond}
Zero-bias conductance [\gzero] calculated using DFT and many-body full-COHSEX and junction coupling energy ($E_\mathrm{coupl}$) for the BDT-h, BDT-p and BDT-n junctions.
The absolute [$\Delta_a \mathcal{G} = \mathcal{G}(\mathrm{DFT}) - \mathcal{G}(\textrm{full-COHSEX})$] and relative [$\Delta_r \mathcal{G} = \Delta_a \mathcal{G} / \mathcal{G}(\mathrm{DFT})$] differences between the DFT and many-body full-COHSEX conductances are also reported.
All conductance values are given in $\go$ units, while the junction coupling energy is given in eV.
For BDT, the widely accepted value of the experimental conductance is 0.010~$\go$, as in Refs.~\cite{xiao_measurement_2004, horiguchi_electron_2009}.
}
\end{table}

%%%%%%%%%%%%%%%%%%%%%%%%%%%%%%%%%%
\subsection{Zero-bias conductance}
%%%%%%%%%%%%%%%%%%%%%%%%%%%%%%%%%%
For the three junctions considered here, the calculated zero-bias conductances \gzero\ using DFT-GGA and many-body full-COHSEX are reported in Table~\ref{table:cond}.
The DFT conductances of 0.305, 0.365, and 0.034~$\go$ for BDT-h, BDT-p and BDT-n respectively, are in agreement with previous results from the literature \cite{ di_ventra_first-principles_2000, stokbro_theoretical_2003, ThygesenJacobsen2005, ning_first-principles_2007, toher_effects_2008, MowbrayThygesen08, StrangeThygesen11, ning_correlation_2014}.
The main results of this work are the full-COHSEX conductances.
For all three bonding geometries, the conductance reduces upon the inclusion of many-body effects. However, we observe a \textit{very different reduction depending on the molecule-lead bonding and the junction geometry}.
While the conduction stays almost at the DFT level for BDT-p (with a negligible reduction of only 0.003~$\go$ and 0.8\% relative to DFT), we observe for BDT-n an absolute reduction of 0.014~$\go$ (corresponding to a $41$\% change with respect to DFT).

This finding leads to several conclusions.
If this reduction of the conductance were solely associated to the opening of the HOMO-LUMO gap of the isolated BDT molecule (possibly reduced by classical screening due to the leads), one should not observe such important variations between different molecule-lead bonding and junction geometries. 
Indeed, the diag-$G_0W_0$ or diag-COHSEX opening of the HOMO-LUMO gap for the isolated (noncontacted) BDT is the same for the different junctions (since they involve the same molecule).
Furthermore, the effect of the classical screening by the leads can only vary by changing the lead-molecule distance and the latter is the same for BDT-p and BDT-n, and slightly shorter for BDT-h.
Therefore, we are led to conclude that the reduction of the conductance upon inclusion of many-body corrections (as induced by a realistic self-energy) strongly depends on the metal-molecule geometry and bonding regime. This is consistent with the presence of molecule-lead hybridization in (some of) the junctions.
A model self-energy, like in the DFT+$\Sigma$ image-charge model, accounting only for the many-body opening of the isolated molecule HOMO-LUMO gap and its reduction due to the classical screening by the leads \cite{ QuekNeaton07, MowbrayThygesen08} \textit{necessarily misses the dependence of the GW correction on the metal-molecule bonding regime} found here.
\textit{The same also holds for an ab initio self-energy restricted to the molecule only,} i.e., neglecting the metal-molecule coupling geometry.
One must take into account the correlations between the molecule and the leads, as well as the many-body modifications to the hybridization of the junction wave functions.
The latter can be expected to depend on the lead-molecule bonding strength and character.

In an effort to understand the relationship between many-body corrections and the lead-molecule coupling, in the last column of Table~\ref{table:cond} we report the metal-molecule binding energy $E_\mathrm{coupl}$.
The latter is defined as half the difference between the junction total energy and the energies of the isolated molecule and leads:
\begin{equation}
E_\mathrm{coupl} = \frac{1}{2}(E_\mathrm{junc} - E_\mathrm{mol} - E_\mathrm{leads}),
\end{equation}
where the factor $1/2$ accounts for the fact that there are two leads.
This quantifies the strength of the metal-molecule bond.
As expected, we can see that BDT-p (in which sulfur is directly bonded to a single gold atom) is the junction presenting the strongest metal-molecule bond, with a coupling energy of $-1.56$~eV, followed by BDT-h (in which the sulfur atom is bonded to three gold atoms).
Finally, BDT-n (in which the sulfur is still bonded to the undissociated hydrogen atom) presents the weakest metal-molecule bond.
Notice that $E_\mathrm{coupl}$ should not be confused with the junction relative stability.
For instance, we have calculated the relative stability of BDT-n with respect to BDT-p, $E_\mathrm{BDT\textrm{-}n} - E_\mathrm{BDT\textrm{-}p} - E_\mathrm{H_2}$, and found a difference of $-0.23$~eV, showing that BDT-n is more stable and confirming previous findings \cite{ning_correlation_2014, zhou_methanethiol_2006, demir_identification_2012}.

In the limits of the reduced sample of geometries for BDT, our study seems to point to a relationship between the reduction of the conductance due to correlations and the lead-molecule coupling strength.
This can be seen when comparing the many-body weight on the conductance $\Delta_r \mathcal{G}=1-\mathcal{G}^\mathrm{MB}/\mathcal{G}^\mathrm{DFT}$ to $E_\mathrm{coupl}$ (fifth and sixth columns of Table~\ref{table:cond}).
The weaker the bond, the larger the many-body reduction.
This implies that many-body corrections should be taken into account for weakly coupled junction in order to obtain more accurate estimates of the conductance.
On the other hand, interestingly our findings suggest that many-body effects can be safely neglected in strongly coupled junctions, DFT being in these cases a good level of approximation for the conductance.

The experimental value of the BDT@Au zero-bias conductance \gzero\ has been a debate over decades.
Since the pioneering work of Reed \textit{et al.}~\cite{reed_conductance_1997}, a broad-range of experimental results have been reported \cite{ bruot_mechanically_2012, kim_benzenedithiol:_2011, ulrich_variability_2006, horiguchi_electron_2009} ranging from $10^{-4}$ up to 0.5~$\mathcal{G}_0$. 
Today, a wide consensus has been achieved on the value of 0.011~$\mathcal{G}_0$ using statistical analysis on a sample of thousands of measurements \cite{ xiao_measurement_2004, horiguchi_electron_2009}.
We compare this experimental value to our theoretical conductance for the BDT-n junction, that has been found to be the most stable geometry in our set.
The computed many-body COHSEX BDT-n conductance, 0.020~$\go$, while still overestimating by a factor 1.8 the experimental consensus value of 0.011~$\go$ \cite{ xiao_measurement_2004, horiguchi_electron_2009} provides however a net improvement over the DFT value of 0.034~$\go$ (which is three times larger than the experiment).

\begin{figure}
\centering
\includegraphics{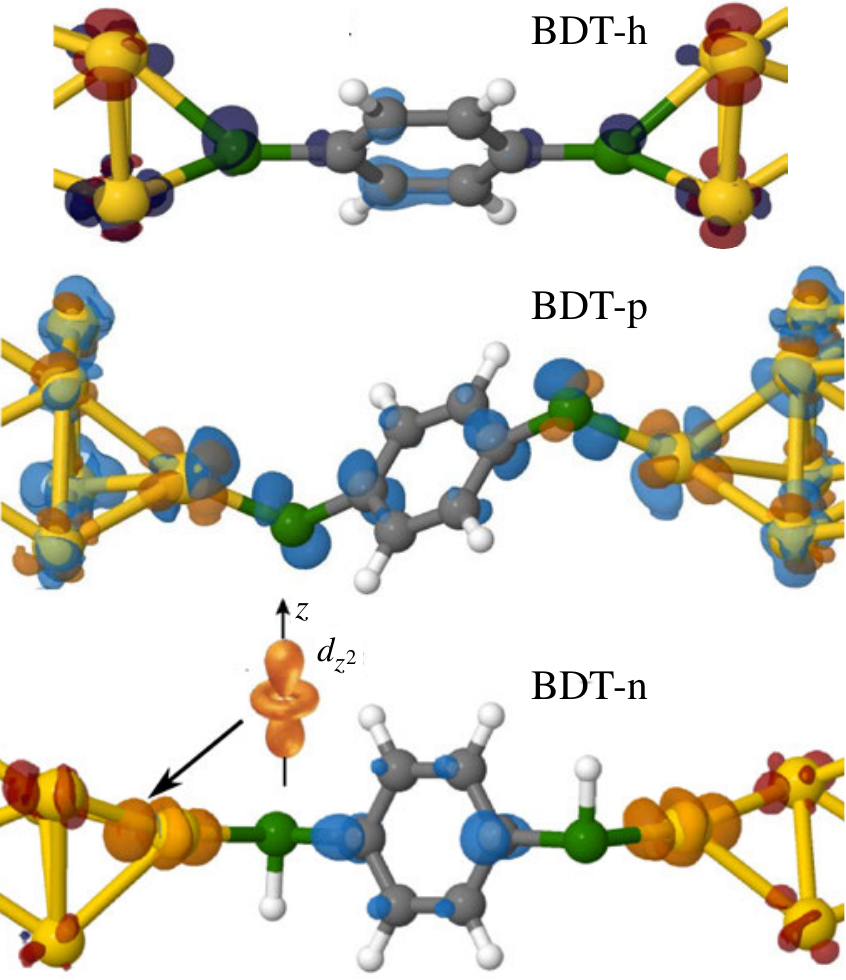}
\caption[LDOS for BDT-h, BDT-p and BDT-n]{
Local density-of-states (LDOS) difference between full-COHSEX and diag-COHSEX calculated in an energy window of 0.8~eV around the Fermi level for BDT-h, BDT-p and BDT-n.
Four isovalues are represented: $+4\rho$ in orange, $+1\rho$ in dark red, $-1\rho$ in dark blue, and $-4\rho$ in light blue, with $5\times10^{-4}$  $e^-$/\AA$^3$ in BDT-h, and $6\times10^{-4}$  $e^-$/\AA$^3$ in BDT-p and BDT-n.
The Au, C, S, and H atoms are represented by yellow, grey, green, and white spheres, respectively.
}
\label{fig:ldos}
\end{figure}

%%%%%%%%%%%%%%%%%%%%%%%%%%%%%%%
\subsection{Local DOS analysis}
\label{sec:ldos}
%%%%%%%%%%%%%%%%%%%%%%%%%%%%%%%
To gain more insight on the changes of the conductance at $E = E_F = 0$, we compute the local density of states~(LDOS) in an energy window $\Delta E = [-0.4,+0.4]$~eV around the Fermi energy $E_F$,
\begin{equation}
\mathrm{LDOS}(\mathbf{r}) = 
  \sum_i
  \int_{\Delta E} d\omega \,
  \delta(\omega-\epsilon_i)
  |\phi_i(\mathbf{r})|^2 .
\end{equation}
The difference in the LDOS between the full- and diag-COHSEX, as plotted in Fig.~\ref{fig:ldos}, directly shows the difference between full-COHSEX and DFT wave functions for all the states around the Fermi energy upon which the zero-bias conductance directly depends. 
The plot shows the modification to the wave functions due to off-diagonal elements of the self-energy matrix which are directly responsible for the changes in the zero-bias conductance (see also Ref.~\cite{RangelRignanese11}).

By comparing the three cases we can try to understand the different behavior of the many-body corrections to the conductance.
The LDOS modification on the benzene ring looks the same in BDT-p and BDT-n (i.e., the two opposite cases with respect to the conductance reduction).
And, there are some similarities also with BDT-h.
This excludes a direct relation between the modifications in the conductance and the wave functions on the molecule.
In contrast, there are clear differences in the LDOS on the sulfur atoms.
Since they are largest in BDT-p and negligible in BDT-n, they also cannot be related to the conductance reduction which is largest in BDT-n.
Hence, the different behavior in the conductance reduction seems to be related to the modification of the wave functions on the gold atoms.
In fact, we find LDOS changes on these atoms of all junctions, which are larger in both BDT-p and BDT-n.

Moreover, we notice different LDOS modifications in BDT-p and BDT-n.
In the latter, we observe an increase in the wave function of $e_g (d_{z^2})$ character, which is anisotropic along the bond axis (see the typical ring shaped lobes on the gold adatom); whereas in BDT-p there is an important but \textit{local} redistribution of charge with a seemingly mixed $t_{2g}$ orbitals isotropic character.
The anisotropic \cite{RangelRignanese11,botti_anisotropy_2002} increase in the $e_g (d_{z^2})$ gold wave functions seems hence to drive the conductance reduction due to correlation effects.

%%%%%%%%%%%%%%%%%%%%%
\section{Conclusions}
%%%%%%%%%%%%%%%%%%%%%
When considering many-body corrections on complete benzene-dithiol~(BDT) gold junctions, we have found very different trends (scanning across a set of different geometries) in the effects on the zero-bias conductance reduction. 
These are largest in weakly coupled molecule-leads junctions and seem associated to an increased anisotropic $e_g (d_{z^2})$ wave function character on gold atoms.
On the other hand, they are much smaller in strongly bond metal-molecule junctions, so that the DFT zero-bias conductance is a good approximation in these systems.
Our findings question the use of many-body corrections that do not depend on the molecule-lead coupling.

%%%%%%%%%%%%%%%%%%%%%%%%%%
\section{Acknowledgments}
%%%%%%%%%%%%%%%%%%%%%%%%%
V.O.~ acknowledges access to the computational resources of the french national GENGI-IDRIS supercomputing centers.
A.F. acknowledges partial support from the EU Centre of Excellence ``MaX'' (MAterials design at the eXascale, Grant No. 676598).
G.-M.R. acknowledges financial support from the F.R.S.-FNRS and access to various computational resources: the Tier-1 supercomputer of the F{\'e}d{\'e}ration Wallonie-Bruxelles funded by the Walloon Region (Grant No. 1117545), and all the facilities provided by the Universit{\'e} Catholique de Louvain (CISM/UCL) and by the Consortium des {\'E}quipements de Calcul Intensif en F{\'e}d{\'e}ration Wallonie Bruxelles (C{\'E}CI).

%%%%%%%%%%%%%%%%%%%%
\bibliography{bdtmbvscoup}
%%%%%%%%%%%%%%%%%%%%

\end{document}